\documentclass[aps,pre,twocolumn,10pt]{revtex4-2}
\usepackage{graphicx}
\usepackage{graphics}
\usepackage{latexsym}
\usepackage{amsmath, amsthm, amssymb, wasysym}
\usepackage{epsfig}
\usepackage[mathscr]{euscript}
\usepackage{blindtext}
\usepackage[normalem]{ulem}
\usepackage{xcolor}
\usepackage{hyperref}  
\usepackage{cleveref}
\usepackage{comment}
\begin{document}
\title{Bunching of extreme events on complex network}

\author{Sarvesh K. Upadhyay$^1$, Vimal Kishore$^1$, Sanjay Kumar$^1$}
\author{R. E. Amritkar$^{1,2}$\thanks{Corresponding author: ravin012@gmail.com}}

\email{ravin012@gmail.com}
\affiliation{$^1$Department of Physics, Banaras Hindu University, Varanasi 221005, India.\\
$^2$Physical Research Laboratory, Ahmedabad 380009, India}

\date{\today}

\begin{abstract}
Extreme events such as earthquakes, floods, and power blackouts often display burst phenomena where multiple extreme events occur in quick succession or in bunches. This study examines bunching of extreme events on a complex network using a random walk transport model. We find that in a modular network, a small cluster sparsely connected with the rest, shows bunching and correlations among extreme events. The bunching and correlations emerge naturally in our system. We use several characterization techniques, namely the recurrence time distribution, autocorrelation function, bursty trains, burstiness parameter and memory coefficient to quantify the bunching and correlations of extreme events. Our study shows that the network structure plays a significant role in the bunching of extreme events.

\end{abstract}
    

\maketitle
\noindent
\section{Introduction}

Extreme events (EEs) are ubiquitous in nature and are observed across a range of fields, spanning from natural disasters and climate anomalies to financial crashes \cite{albeverio2006extreme,bak,PhysRevLett.88.178501,tokyo.ac.jp:00037571,Scholz2019}.

 A key phenomenon associated with  EEs is their tendency to occur in bunches \footnote{The common terminology to define a group of EEs is a `cluster'. However, we use the term cluster to define a group of nodes and hence use the term `bunch' to address a group of EEs.} within a limited region and time as observed with the aftershocks \cite{Pollitz2012-ck} and foreshocks \cite{10.1785/0220220397} that accompany a major earthquake. Such bunches of EEs typically exhibit bursts of intense activity separated by extended periods of inactivity \cite{Barabasi2005,Oliveira2005}.  Other examples of bunching of EEs include water levels of the Nile near Cairo \cite{Giuliano}, six floods in three years in Harbertonford, UK \cite{doi:10.1680/muen.2005.158.2.107}, extreme precipitation events in Finland  \cite{https://doi.org/10.1002/joc.5867} and southern Switzerland \cite{ClusteringofRegionalScaleExtremePrecipitationEventsinSouthernSwitzerland}, northern hemisphere temperature spikes \cite{cp-8-227-2012}, road accident bunches in Tiruvananthapuram, India \cite{PRASANNAKUMAR2011317}, and the prolonged series of bank failures starting in 2008 \cite{IVASHINA2010319}, sunspots \cite{Wheatland_1998}, and neuronal activity \cite{Kemuriyama2010-vc,Grace1984-xs}, traffic jams \cite{Treiber2013}.

The bunching of EEs indicates some correlations among these events. Theoretical attempts to understand the EEs mostly consists of generating correlated data using various techniques such as the Fourier-filtering \cite{PhysRevE.53.5445}, memory-based dynamics \cite{Karsai2012}, fractional Ornstein - Uhlenbeck equation \cite{Telesca2020-zg} and the local non-stationarity in the underlying processes  \cite{towe10.1007}.
There is no known unique characterization technique for studying the bunching of EEs. Hence, one uses several characterization techniques \cite{Jo2023}. A commonly used technique is the recurrence time distribution \cite{PhysRevLett.94.048701,PhysRevLett.92.108501}. An exponential distribution typically indicates an independent and identically distributed (iid) sequence of events. Deviations from this exponential pattern suggest correlations among events and hence bunching of events. Other techniques include the autocorrelation function, bursty trains and burst size distribution, memory coefficient, and burstiness parameter \cite{Karsai2018}.

Several studies have examined EEs in dynamical systems and random walkers \cite{NAGCHOWDHURY20221,PhysRevE.97.062102}. Many complex dynamical systems can be usefully modeled using complex networks \cite{RevModPhys.74.47,doi:10.1137/S003614450342480}.
Kishore, Santhanam and Amritkar \cite{PhysRevLett.106.188701,PhysRevE.85.056120} have studied EEs on nodes of a network using random walk model and find that the probability of EEs decreases with the degree of a node.  
Here, we use this model of random walks on a network, to investigate the bunching of EEs in networks. We observe bunching of EEs in modular networks and find that network structure plays an important role in the bunching of EEs.

\section{Model and Method}

Following Ref.~\cite{PhysRevLett.106.188701}, we start with a connected, undirected network of $N$ nodes and $E$ edges and put a random walker on a node of the network. At each time step, the walker takes a step with equal probability to one of the neighboring nodes through an edge. Asymptotically the random walker has a stationary distribution given by~\cite{PhysRevLett.92.118701}
\begin{align}
p = p_0(K) = \frac{K}{2E}, \label{p-asy}
\end{align} 
where $p_0(K)$ is the stationary probability that a walker is on a node with degree K.

We now put $W$ independent non-interacting walkers on the network. The probability that there are $w$ walkers on a node with degree $K$ is a binomial distribution,
\begin{align}
    f(w) = \binom{W}{w} p^w (1-p)^{W-w}.\label{binom}
\end{align}
The mean number of walkers on a node and the standard deviation are $\langle w \rangle = Wp$, $\sigma^2 = Wp(1 - p)$. 
We now define an extreme event on a node as the number of walkers on that node exceeding a threshold given by  \cite{PhysRevLett.106.188701}
\begin{align}
    q = \langle w \rangle + m \sigma  \label{threshold} 
\end{align} 
where the threshold parameter $m$ is a real number. Thus, the EE probability for a node with degree $K$ is 
 \begin{align}
     F(K) =\sum_{w= \lfloor q \rfloor +1}^{W} f(w) = I_p(\lfloor q \rfloor +1,W-\lfloor q \rfloor ) \label{Fk}
 \end{align}
 where $I_p(.,.)$ is the regularized incomplete beta function and $\lfloor q \rfloor$ is the largest integer less than $q$ \cite{PhysRevLett.106.188701}.

\section{Bunching of extreme events}

\begin{figure}
   \includegraphics[scale=0.6,clip]{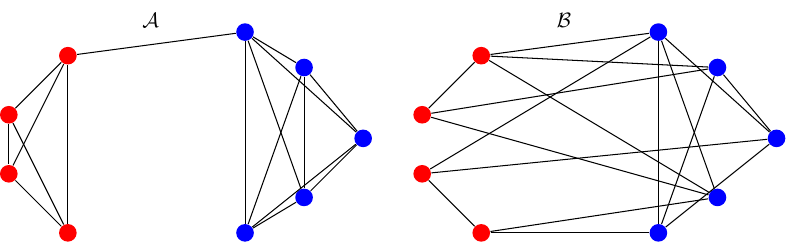}
  \caption{The figure shows an illustration of the two types of networks $\cal{A}$ and $\cal{B}$ used in the text. The network $\cal{A}$($n_1-n_2$) with 9 nodes, is divided into two
clusters of $n_1=4$ and $n_2=5$ nodes such that each cluster is completely connected and the two clusters are connected by a single bridge. Network $\cal{B}$ has the same number of nodes with the same degrees as in $\cal{A}$, but with randomly connected edges.}  
  \label{fignetAB}
\end{figure}

Though Eq.~(\ref{p-asy}) gives the asymptotic stationary distribution, the number of walkers on a node fluctuates according to Eq.~(\ref{binom}). When the fluctuations are large we get EEs and their probability is given by Eq.~(\ref{Fk}). In general, the fluctuations and hence the EEs are not correlated. However, we find that
the structure of a network can play an important role in developing correlations. To see this behavior we consider two different networks $\cal{A}$ and $\cal{B}$ as illustrated in Fig.~\ref{fignetAB}.

\subsection{Network \texorpdfstring{$\cal{A}$}: Network with a small and a large cluster}

\begin{figure}
  \includegraphics[width=1\linewidth]{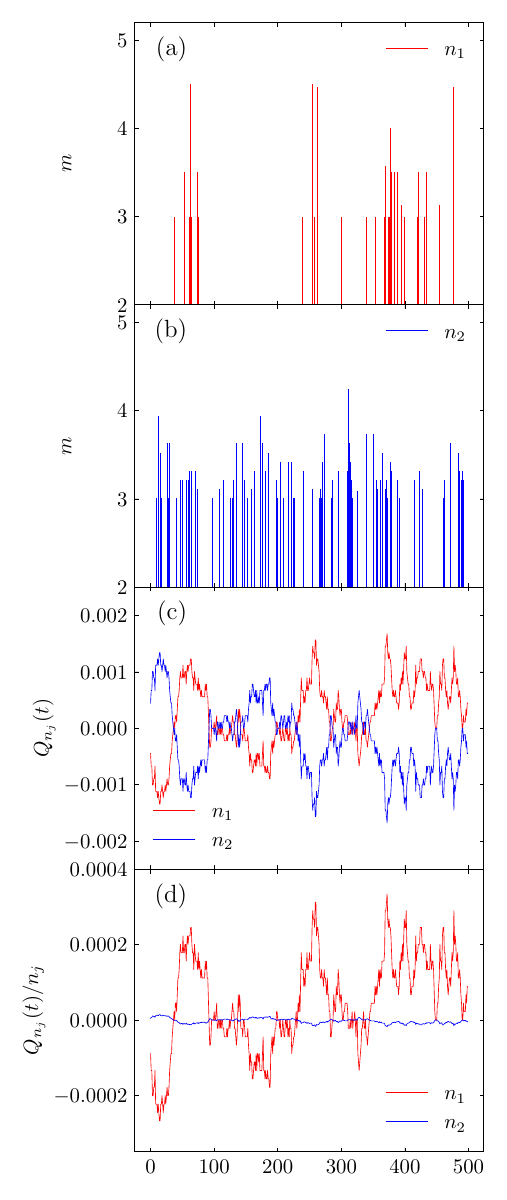}
   \caption{\label{figAexev}
   (a) and (b) The EEs of magnitude greater than $3\sigma$ for a time window of 500 units, are shown for the two clusters $n_1=5$ (red online) and $n_2=95$
 (blue online)  respectively for  network $\cal{A}$. Here and in subsequent analysis we use $W=2E$. (c) and (d) The  probability difference $Q_{n_j}(t)$ (Eq.~(\ref{Qjt})) and the probability difference per node $Q_{n_j}(t)/n_j$ are respectively plotted as a function of $t$ for the two clusters $n_1$ and $n_2$.}

\end{figure}

We start with a network of type $\cal{A}$ with two clusters $n_1=5, \ n_2=95$, a small and a large cluster. Fig.~\ref{figAexev}(a) and (b) show EEs of magnitude larger than $3\sigma$ in clusters $n_1$ and $n_2$ respectively. Though both clusters show EEs, those in the small cluster show bunching.
For the small cluster, from $t=54$ to 76, we see a bunching with one EE of magnitude greater than $4.5\sigma$ ($t=63$) and several smaller magnitude EEs while from $t=239$ to 263 we see two large EEs ($t=255$ and $263$) with other smaller magnitude events. From $t= 339$ to 477, we see several EEs with one large EE forming a miniswarm \footnote{Similar structures such as doublet \cite{RUIZ2017102,Weber} and swarms 
\cite{doi:10.1126/sciadv.ado1469,Horalek2015} are reported in earthquake data.}. No such clear bunching of EEs is observed in the large cluster $n_2=95$ \footnote{This will be further clarified when we consider characterization techniques. Also, The larger cluster has more EEs since the number of nodes is large. Actually, per node EEs in the large cluster is smaller than that in the small cluster due to the larger degree \cite{PhysRevLett.106.188701}.}.

To understand the bunching in small cluster $n_1$, consider the observed probability of finding a walker in the $j^{th}$ cluster at time $t$,
$P(t,n_j)=\sum_{i \in n_j} w_i(t)/W$. We define 
the probability deviation $Q_{n_j}(t)$ as the difference between the observed and the stationary probability of a walker to be in the $j^{th}$ cluster at time $t$.
\begin{align}
    Q_{n_j}(t) =P(t,n_j) - P_0(n_j) =\sum_{i \in n_j} \left(\frac{w_i}{W} -\frac{K_i}{2E} \right)
    \label{Qjt}
\end{align}

\begin{figure}   
 \includegraphics[width=1\linewidth]{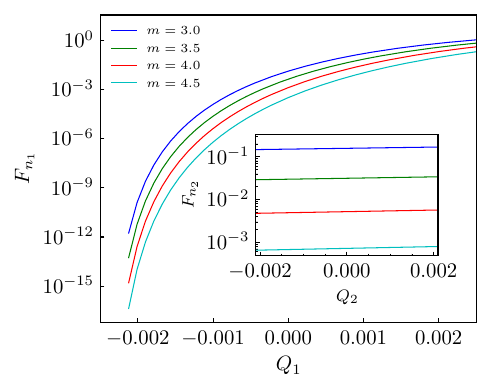}
   \caption{\label{figFjQ}
   Probability of EEs for the small cluster, $F_{n_1}$, as a function of the probability difference $Q_{n_1}$ for different $m$ (Eq.~(\ref{Qjt})). The inset shows the same for the large cluster.}
\end{figure}

In Fig.~\ref{figAexev}(c) we plot the probability difference $Q_{n_j}(t)$ vs $t$ for the two clusters $n_1$ and $n_2$. We see that for the small cluster $n_1$, most of the EEs occur when $Q_{n_j}(t)$ is positive and very few EEs occur when $Q_{n_1}(t)$ is negative. This explains the bunching effect which occurs when $Q_{n_j}(t)$ remains positive for a reasonable duration of time.

The reason for this, can be understood from Fig.~\ref{figFjQ} where we plot $F_{n_j} = \sum_{i\in n_j} F(K_i)$, the probability of EEs on a cluster $n_j$ as a function of $Q$. For the small network (Fig.~\ref{figFjQ}), the probability increases almost exponentially for $Q>0$ and falls sharply (better than exponential) for $Q<0$.

On the other hand, for the large cluster $n_2$, there is no correlation between the sign of $Q_{n_j}(t)$ and the EEs (Figs.~\ref{figAexev}(b) and (c)).  The reasons are as follows. Firstly, from Fig.~\ref{figAexev}(d) where we plot the probability difference per node $Q_{n_j}(t)/n_j$ vs $t$ for the two clusters $n_1$ and $n_2$, we see that probability difference per node remains significant for the small cluster while it becomes negligible for the large cluster. Secondly, for the large cluster the probability of EEs is nearly constant as a function of $Q$ (Fig.~\ref{figFjQ}(Inset)).

\subsection{Network \texorpdfstring{$\cal{B}$}: Random network}

\begin{figure}
 \includegraphics[width=1\linewidth]{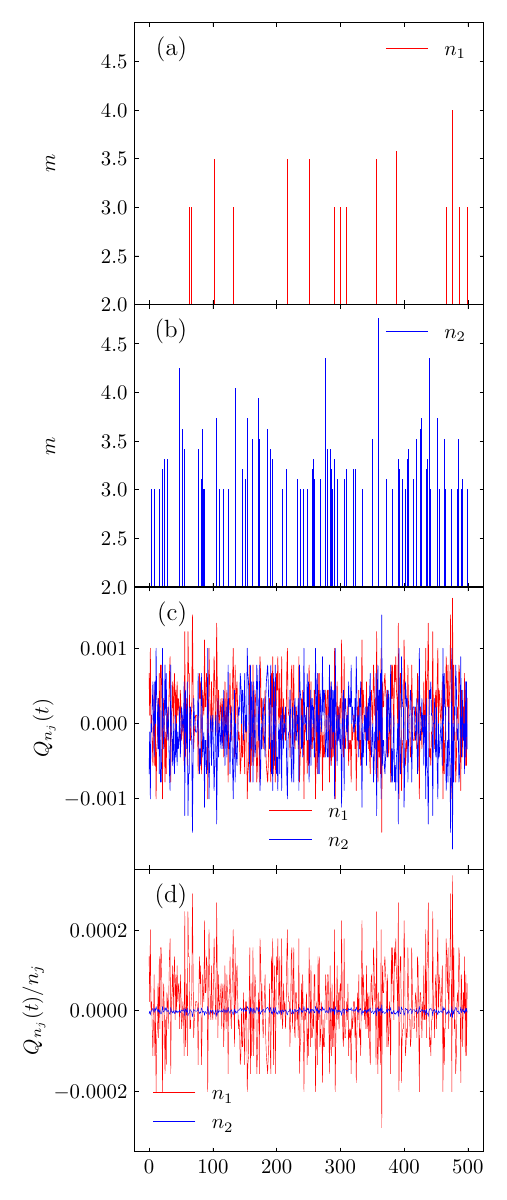} 
 \caption{\label{fig:rand_net} (a) and (b) The EEs of magnitude greater than $3\sigma$ for a time window of 500 units, are shown for the two clusters $n_1=5$ and $n_2=95$ respectively in a network of type $\cal{B}$. (c) and (d) plot respectively, the  probability difference $Q_{n_j}(t)$ (Eq.~(\ref{Qjt})) and the probability difference per node  $Q_{n_j}(t)/n_j$ as a function of $t$ for the two clusters $n_1$ (red) and $n_2$ (blue).}
\end{figure}

We now consider the network $\cal{B}$ in Fig.~\ref{fignetAB}. It has the same number of nodes and same degrees as in network $\cal{A}$, but the edges are random. We can artificially split it into two clusters by grouping the nodes with degrees similar to network $\cal{A}$. Thus, we split a network with $N=100$ into two clusters $n_1=5$ and $n_2=95$ with the same degrees as in network $\cal{A}$(5-95) but with random connections. Fig.~\ref{fig:rand_net} is a repeat of Fig.~\ref{figAexev} for network $\cal{A}$, but now for network $\cal{B}$. 
The EEs do not show bunching effect (Figs.~\ref{fig:rand_net}(a) and~(b)) and follow iid EE properties for both clusters $n_1$ and $n_2$. 
Fig.~\ref{fig:rand_net}(c) shows a random behavior for the probability difference $Q_{n_j}(t)$ and no oscillatory behavior is seen. 
In Fig.~\ref{fig:rand_net}(d), we see that the probability difference per node remains significant for the small cluster while it becomes negligible for the large cluster as in Fig.~\ref{figAexev}(d). 

The main difference between the dynamics of networks $\cal{A}$ and $\cal{B}$ is that in network $\cal{A}$ walkers are trapped in the two clusters due to the single bridge connecting the two clusters $n_1$ and $n_2$. This leads to the oscillations of the number of walkers in the two clusters. No such trapping of walkers occurs in network $\cal{B}$ since the edges are random and, hence, no oscillations  are seen. 

\subsection{Oscillations of walkers in network \texorpdfstring{$\cal{A}$}.}
\begin{figure}
  
  \includegraphics[width=1\linewidth]{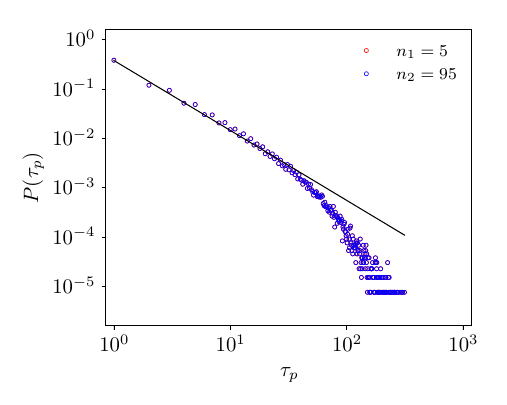}
   \caption{\label{figtaup} The distribution $p(\tau_p)$ for the two clusters $n_1=5$ and $n_2=95$ in a modular network $\cal{A}$($5-95$) as a function of the time period $\tau_p$ for positive values of $Q_{n_j}=P(t,n_j) - P_0(n_j)$. The observed values show a power law behavior upto $\tau_p \simeq 40$ (black lines) $P(\tau_p) \sim A\tau_p^{-B}$ where $A,B= 0.221,1.057$ . Since $Q_{n_1}+Q_{n_2}=0$, the plot of $P(\tau_p)$ for negative values of $Q$ is the same with symbol colors interchanged.}
\end{figure}
From Fig.~\ref{figAexev}(c), we see that the probability difference $Q_{n_j}(t)$ for a cluster in network $\cal{A}$, i.e. the excess probability of a walker being in a cluster, shows oscillatory behavior. These oscillations arise because the walkers get trapped in a cluster due to the single bridge connecting the two clusters. As discussed above, these oscillations are primarily responsible for the bunching of EEs in small cluster which occur when $Q$ is positive \footnote{We can compare this with the physics of earthquakes. When a bunch of earthquakes is observed in a region, the average geothermal energy in that region is larger than the quiescent value.}. We now consider some properties of these oscillations. 
\begin{figure}
   \includegraphics[width=1\linewidth]{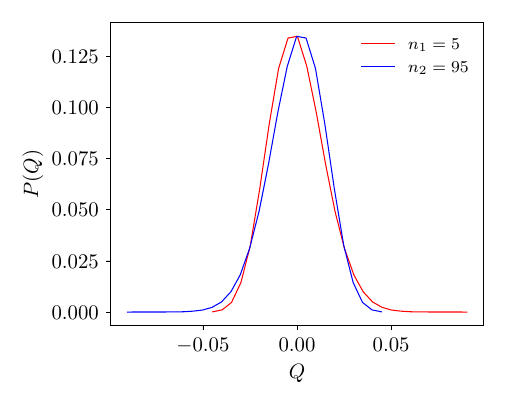}
 \caption{ \label{figtaup1}The distribution $P(Q)$ for the two clusters $n_1=5$ and $n_2=95$ in a modular network $\cal{A}$($5-95$) as a function of $Q$.} 
\end{figure}
The oscillations of number of walkers in the two clusters of network $\cal{A}$ can be identified by positive and negative values of the probability difference $Q_{n_j}(t)$ as seen in Fig.~\ref{figAexev}(c). Let $\tau_p$ be the time interval or period for which $Q_{n_j}$ has the same sign. In Fig.~\ref{figtaup}, we plot, the probability distribution of $P(\tau_p)$ as a function of $\tau_p$ for positive values of $Q_{n_j}$ for the two clusters $n_1=5$ and $n_2=95$. Since, $Q_1+Q_2=0$, the distribution for the negative values is the same with the colors and symbols interchanged. For $\tau_p<\tau_p^c$, where $\tau_p^c\sim 40$, $p(\tau_p)$ shows a power law behavior while for larger values of $\tau_p$, $P(\tau_p)$ falls sharply.

Fig.~\ref{figtaup1} shows the probability distribution of $P(Q_{n_j})$ as a function of $Q$ for the two clusters $n_1=5, \ \ n_2=95$. Though the peak for $P(Q_{n_1})$ occurs for a smaller $Q$ value than for $P(Q_{n_2})$, $P(Q_{n_1})$ has a fatter tail for positive $Q$. However, these differences are small and two distributions are almost the same.

\section{Characterization of bunching}

The above analysis tells us that EE sequences in the small cluster exhibit bunching of EEs and hence must have temporal correlations.

\subsection{Recurrence time}

Let $\tau_r$ denote the recurrence time of EEs, also called the inter event time. 
If we assume that the EEs follow iid (independent and identically distributed), then the probability distribution of recurrence time for cluster $n_j$ is given by
\begin{align}
p^{iid}(\tau_r) = F_{n_j} (1-F_{n_j})^{\tau_r-1}, \; \; \tau_r=1,2,3,\cdots \label{ptaur}
\end{align}
The mean recurrence time and the standard deviation are $\langle \tau_r^{iid} \rangle = 1/F_{n_j}$ and $\sigma_r^{iid} = \sqrt{1-F_{n_j}}/F_{n_j}$.
Since $F_{n_j}$ is typically small, $\sigma_r^{iid} \approx \langle \tau_r^{iid} \rangle$.

\begin{figure}
\includegraphics[width=1\linewidth]{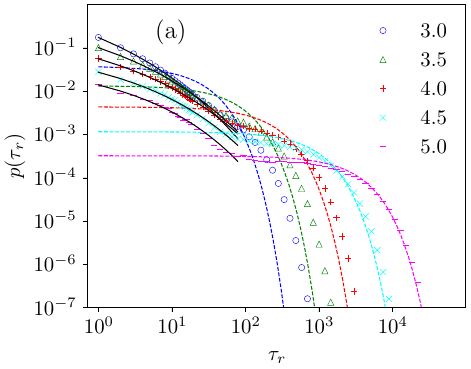}
 \includegraphics[width=1\linewidth]{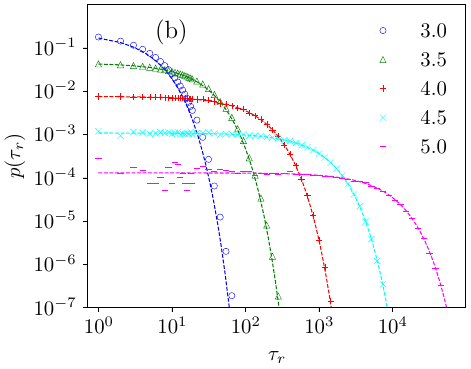}
  \caption{\label{figptaur}(a) Observed probability distribution $p(\tau_r)$ of recurrence time $\tau_r$ for the small cluster $n_1=5$ (network ${\cal{A}}(5-95)$), for five different values of $m$. For $\tau_r<=20$ we use linear binning with all $\tau_r$ values while for larger $\tau_r$ values we use logarithmic binning,  $(a^n,a^{n+1})$ with $a=1.2$. The observed probability distribution shows considerable deviation from that for the iid EEs shown by the dashed lines (colored online), Eq.~(\ref{ptaur}). The black solid lines show a stretched exponential behavior $p(\tau_r) = C_re^{-D_r\tau_r^{\gamma_r}}$. The parameters are given in Table~I, Appendix. (b) Similar plot for probability distribution $p(\tau_r)$ as in (a) for the cluster $n_2=95$. The observed probability distribution shows good agreement with the theoretical prediction of Eq.~(\ref{ptaur}) for iid EEs (dashed lines - color online).}
\end{figure}

\begin{figure}
 \includegraphics[width=1\linewidth]{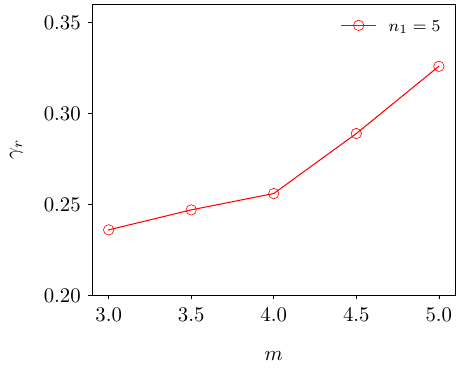} 
  \caption{ The figure plots the stretched exponential parameter $\gamma_r$ for the recurrence time distribution ($p(\tau_r)=C_re^{-D_r\tau_r^{\gamma_r}}$) vs the threshold parameter $m$ for the small cluster $n_1=5$.}
  \label{gammar_ma}
  \end{figure}

In Fig.~\ref{figptaur}(a) we plot the observed probability distribution $p(\tau_r)$ of
recurrence time $\tau_r$ for the small cluster $n_1=5$ for five values of $m$. The dashed lines show
the iid predictions for EEs (Eq. (\ref{ptaur})). We observe
considerable deviations from the iid prediction indicating emergence of correlations among the EEs. For smaller values of $\tau_r$ we observe a stretched exponential behavior
$p(\tau_r) = C_re^{-D_r\tau_r^{\gamma_r}}$. The stretched exponential behavior is roughly observed till some critical value $\tau_r^c$. For $\tau_r \ge \tau_r^c$, $p(\tau_r)$ shows an exponential behavior similar to that of an iid event. The exponent $\gamma_r$ as a function of $m$ is shown in Fig.~\ref{gammar_ma} and the exponent $\gamma_r$ is found to increase with $m$.

    \begin{figure}
\includegraphics[width=1\linewidth]{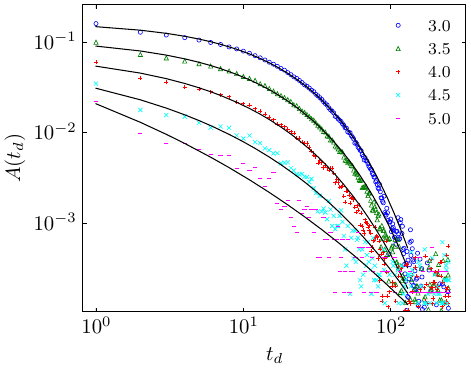}
  \caption{\label{fig_auto} Auto correlation function $A(t_d)$ of the EE sequence with respect to time lag $t_d$ for the small cluster $n_1=5$ (network ${\cal{A}}(5-95)$) for five values of $m$. The black lines show stretched exponential behavior $A(t_d) \sim C_de^{-D_dt_d^{\gamma_d}}$. Parameters of the stretched exponential fit are given in Table~II, Appendix.}
\end{figure}

\begin{figure}
 \includegraphics[width=1\linewidth]{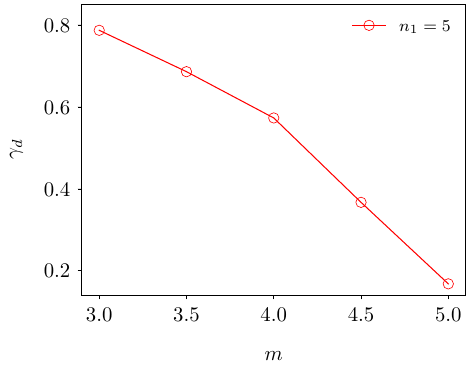} 
  \caption{ Stretched exponential parameter $\gamma_d$ for the autocorrelation function ($p(\tau_d)=C_de^{-D_d\tau_d^{\gamma_d}}$) vs the threshold parameter $m$ for the small cluster $n_1$.}
  \label{gammar_mb}
  \end{figure}

We note that the stretched exponential behavior of recurrence time distribution is reported for observations of several real systems \cite{PhysRevE.73.026117,PhysRevLett.94.048701} and also for simulated data \cite{BUNDE20031,PhysRevE.71.056106,PhysRevE.75.011128}. 
As noted in the discussion of Fig.~\ref{figAexev} the big cluster $n_2=95$ of the modular network $\cal{A}$, does not show any pronounced bunching effect. Hence, all the characterization techniques are expected to show iid EE properties when applied to the cluster $n_2$.

In Figs.~\ref{figptaur}(b), we show the probability distribution $p(\tau_r)$ of recurrence time $\tau_r$ for cluster $n_2$ for different values of $m$. We find that our simulated distribution $p(\tau_r)$ is in good agreement with the theoretical prediction for iid events (Eq.(\ref{ptaur})).

\subsection{Autocorrelation}

The autocorrelation function $A(t_d)$ for delay time or time lag $t_d$, is defined as
\begin{equation}
    A(t_d) = \frac{\langle x(t)x(t + t_d) \rangle - \langle x(t) \rangle \langle x(t+t_d) \rangle} {\sqrt{{(\langle x(t)^2  \rangle -\langle x(t) \rangle ^2)( \langle x(t+t_d)^2\rangle -\langle x(t+t_d) \rangle ^2)}}}
\label{eq:autocorrelation_function}
\end{equation}
where $x(t)$ is 1 if an extreme event occurs at time $t$ and 0 otherwise.

Fig.~\ref{fig_auto} plots the autocorrelation function $A(t_d)$ vs the time lag $t_d$ for the small cluster $n_1=5$ (network $\cal{A}$($5-95$)). The observed values show a stretched exponential behavior (black lines) $A(t_d) \sim C_de^{-D_dt_d^{\gamma_d}}$. The exponent $\gamma_d$ as a function of $m$ is shown in Fig.~\ref{gammar_mb} and it decreases with increasing $m$. 

For the large cluster $n_2=95$, the autocorrelation function is zero within numerical errors as expected.

\subsection{Bursty Train Analysis}

The correlations between recurrence times can also be quantified using the bursty trains. A bursty train refers to a sequence of successive events where the time interval between any two consecutive occurrences is smaller than or equal to a time window, $\Delta t$. The number of events in a bursty train is called the burst size represented by the symbol $b$. It is easy to show that the total number of bursty trains, $N_B(\Delta t)$ is given by,
\begin{align}
N_B(\Delta t) = |\{\tau_i \,|\, \tau_i > \Delta t\}| + 1
\end{align}
where $\tau_i, i=1,2,\ldots$ are the recurrence times. For iid EEs,
\begin{align}
N_B^{iid}(\Delta t) = N_r (1-F_{n_j})^{\Delta t} +1  \label{B_equation}
\end{align}
where $N_r$ is the total number of recurrence times in a given time series.
Also, for the iid EEs, the Burst size distribution, $P_{\Delta t}(b)$, is 
\begin{equation}
P_{\Delta t}^{iid}(b) = (1- (1-F_{n_j})^{\Delta t})^{b-1} (1-F_{n_j})^{\Delta t} .\label{b_equation}
\end{equation}

\begin{figure}

 \includegraphics[width=1\linewidth]{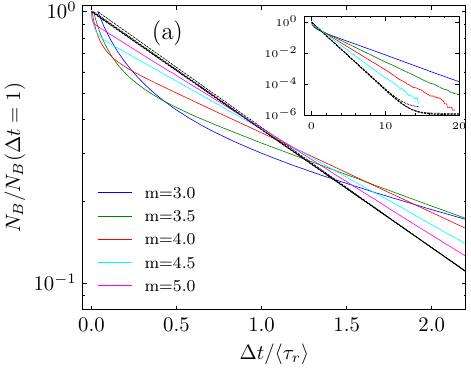}
  \includegraphics[width=1\linewidth]{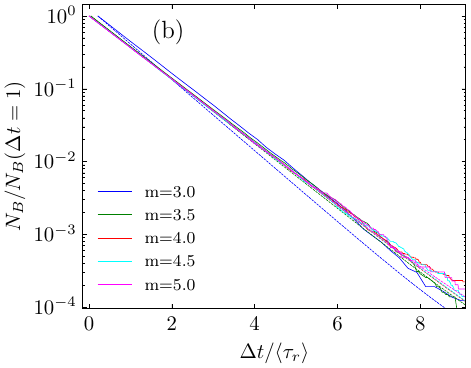}
\caption{\label{figBtrains}(a) Number of burst trains, $N_B(\Delta t)/N_B(1)$ (normalized by $N_B(1)$),  as a function of the scaled time window $\Delta t/ \langle {\tau_r} \rangle$ for the $n_1=5$ cluster of network $\cal{A}$, for five values of $m$, solid lines (colored online). Here, $\langle{\tau_r}\rangle$ is the mean recurrence time of EEs (see Table~IV, Suppl. material). The dashed lines (colored online) are for the iid extreme events (Eq.~(\ref{B_equation})).  The inset shows the behavior for a large times with values upto $\Delta t/\langle {\tau_r}\rangle = 20$. (b) The same plot as in (a), for the large network $n_2$.}

\end{figure}

\begin{figure}
\includegraphics[width=1\linewidth]{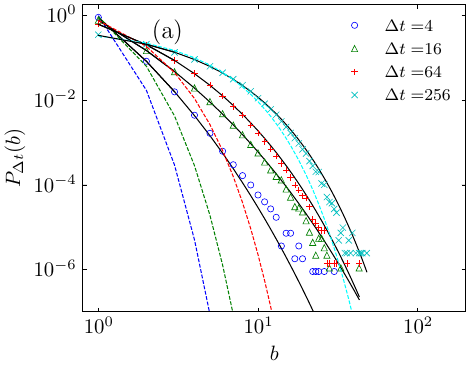}
    \includegraphics[width=1\linewidth]{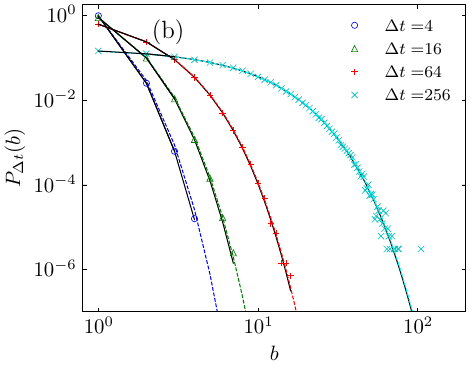}
   \caption{\label{figBtrains1}(a) Burst size  distribution $P_{\Delta t}(b)$ as a function of the bursty train size $b$, for different time windows ($\Delta t$) for the cluster $n_1=5$ for $m=4$. The dashed lines (colored online) show iid behavior (Eq.~(\ref{b_equation})). The black solid lines are fitted stretched exponential form $P_{\Delta t}(b)=C_be^{-D_bb^{\gamma_b}}$. The parameters are listed in Table~III, Appendix. (b) The same plot as in (a) for the large cluster $n_2$.} 
\end{figure}

Fig.~\ref{figBtrains}(a) shows the number of bursty trains, $N_B(\Delta t)/N_B(1)$, as a function of the time window $\Delta t/ \overline{\tau_r}$ for the small cluster $n_1=5$ ($\cal{A}$($5:95$)). The black dotted lines are for the iid EEs.  We see that $N_B(\Delta t)$ is smaller than iid values for small values of $\Delta t$. Since the number of recurrence times, $N_r$, is fixed, a smaller value of $N_B$ indicates larger sizes of bursty trains. This indicates larger correlations for small $\Delta t$ values. We call this as a bunching effect. As $\Delta t$ increases there is a crossover to $N_b$ values larger than iid, indicating smaller sizes of bursty trains. We call this as an anti-bunching effect.

Fig.~\ref{figBtrains}(b) shows the number of bursty trains, $N_B(\Delta t)/N_B(1)$, as a function of the time window $\Delta t/ \overline{\tau_r}$ for the large cluster $n_2=95$. As expected, the data
closely aligns with the iid EEs, Eq. (9).

In Fig. \ref{figBtrains1}(a), we plot the burst size distribution $P_{\Delta t}(b)$ as a function of the bursty train size $b$, for different time windows $\Delta t$ for the small cluster $n_1=5$ ($\cal{A}$($5:95$)) using a threshold of $m=4.0$. The dashed colored curves are for iid EEs which show an exponential behavior (Eq.~(\ref{b_equation})). The observed values show deviations from the exponential behavior, indicating correlations between events. The observed curves are fitted with a stretched exponential function, $P_{\Delta t}(b)=C_be^{-D_bb^{\gamma_b}}$.

The corresponding plot of the burst size distribution $P_{\Delta t}(b)$ as a function of the bursty train size $b$ for the large cluster $n_2=95$ is shown in Fig. \ref{figBtrains1}(b). Again as expected the data closely aligns with the iid EEs, Eq.~(\ref{b_equation}).

\subsection{Burstiness Parameter and Memory Coefficient}

The burstiness parameter $B$ is defined as \cite{Goh_2008}

\begin{align}
    B = \frac{(C_v  - 1)}  {(C_v +1)} =\frac{\sigma_r - \langle{\tau_r}\rangle} {\sigma_r + \langle{\tau_r}\rangle} . 
    \label{bustiness}
\end{align} 
where $\langle{\tau_r}\rangle$ and $\sigma_r$ are the mean and the standard deviation of the recurrence time $\tau_r$ and $C_v=\sigma_r/\langle{\tau_r\rangle}$ is the coefficient of variation.
The memory coefficient is defined as
\begin{align}
M \equiv \frac{1}{n - 1} \sum_{i=1}^{n-1} \frac{(\tau_i - \langle \tau_1 \rangle)(\tau_{i+1} - \langle \tau_2 \rangle)}{\sigma_1 \sigma_2},
\label{memmory coeficient}
\end{align} 
where $<\tau_1>$ ( $<\tau_2>$ ) and $\sigma_1$ ($\sigma_2$ ) are the average and the standard deviation of the
first (last) $N_r  - 1$ recurrence times. For iid EEs the burtiness parameter and memory coefficient are
\begin{align*}
B^{iid} & = \frac{\sqrt{1-F_{n_j}} -1}{\sqrt{1-F_{n_j}}+1} \approx -\frac{1}{4} \ F_{n_j}, \nonumber \\ M^{iid} & = 0 
\label{iid-BM}
\end{align*}

  \begin{figure}
 \includegraphics[width=1\linewidth]{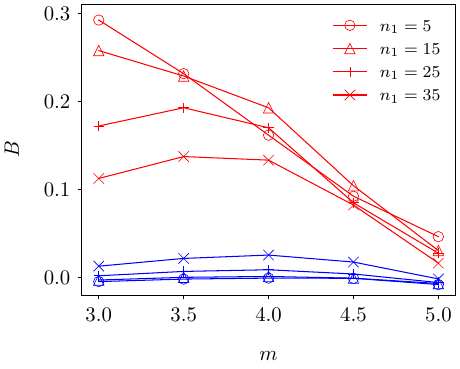}
  \caption{ The variation of (a) the burstiness parameter $B$ with the threshold parameter $m$ for the small cluster (red) and the large cluster (blue) for different values of $n_1$ and the corresponding values of $n_2=100-n_1$.}
  \label{MB_ma}
\end{figure}

  \begin{figure}
 \includegraphics[width=1\linewidth]{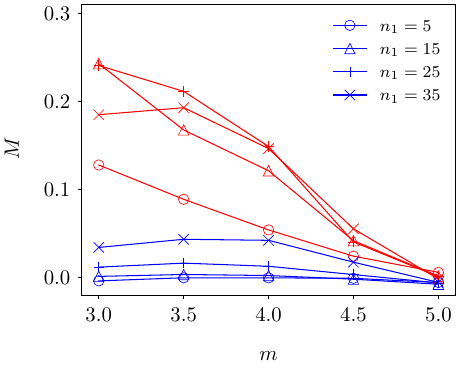}
  \caption{The variation of memory coefficient $M$ with the threshold parameter $m$ for the small cluster (red) and the large cluster (blue) for different values of $n_1$ and the corresponding values of $n_2=100-n_1$.}
  \label{MB_mb}
\end{figure}
In Fig. \ref{MB_ma}, we plot the behavior of the burstiness parameter $B$, for the small cluster (red online) and the large cluster (blue online, with the same symbols as for $n_1$) for different values of $n_1=5,15,25,35$ and $n_2=100-n_1$ as a function of the threshold parameter $m$. The burstiness parameter decreases with $m$ for $n_1=5,15$ while it first increases and then decreases for $n_1=25,35$. For the large cluster, the burstiness parameter remains close to zero for all $m$ values with a slight increase as $n_2$ decreases.

Similarly, Fig. \ref{MB_mb} shows the memory coefficient as a function of $m$. For the small cluster, the memory coefficient decreases as $m$ increases, while for the larger cluster, it stays close to zero with a slight increase as $n_2$ decreases.

\begin{figure}
\centering
\includegraphics[width=1\linewidth]{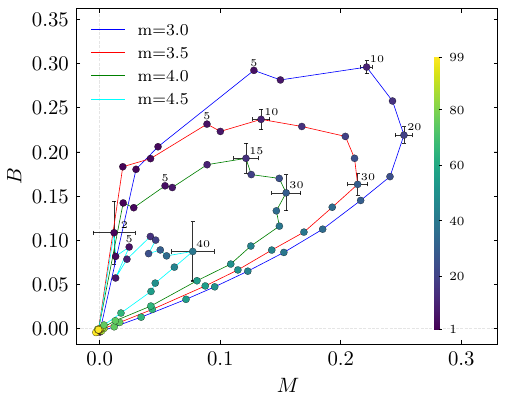}
\caption{\label{figMBa} The $M-B$ phase diagram for different values of $m$. (a) shows the phase diagram with a constant total number of nodes ($N=100$) but varying cluster nodes $n_1=1,2,\ldots 99$ and $n_1+n_2=N$. Color code (online) for the points is used to specify the number of nodes in the cluster $n_1$.
}
\end{figure}
\begin{figure}
 \centering
\includegraphics[width=1\linewidth]{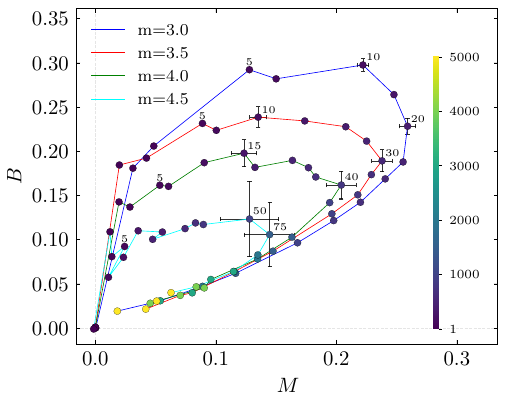}
    \caption{The $M-B$ phase diagram for a constant cluster node ratio ($n_1/N = 1/20$), with varying  $N$. Here the color code specifies the total number of nodes $N$.}
    \label{figMBb}
\end{figure}

However, it is more interesting to discuss 
the $M-B$ phase diagram. We study the $M-B$ phase diagram for different network parameters,  highlighting the
relationship between the memory coefficient $M$, and burstiness
parameter $B$ as network parameters change.

\begin{itemize}
\item[(a)] Constant $N=n_1+n_2$, varying $n_1$: Fig.~\ref{figMBa} plots the $M-B$ phase diagram for cluster $n_1$, for different values of $n_1=1,2,\ldots,99$ and constant $N=100=n_1+n_2$.
We observe that both $M$ and $B$ initially increase as the size of the cluster $n_1$ increases. As the size of the cluster $n_1$ increases further, both $M$ and $B$ start to decrease with $B$ starting to decrease before $M$. The turning point depends on parameter $m$.

It is interesting to see that there is no bunching for the smallest cluster of one node. The reason is that there is no trapping of walkers for a one node cluster and the oscillations observed in Fig.~\ref{figAexev}(c) are absent. As the size of the cluster increases we observe bunching. With further increase in the size, bunching again decreases since the per node effect goes down (Fig.~\ref{figAexev}(d)) and also the probability of EEs remains almost constant as the probability difference $Q$ varies (Fig.~\ref{figFjQ}(inset)).

\item[(b)] Constant ratio $n_1/N$, varying $N$: Fig.~\ref{figMBb} plots the $M-B$ phase diagram for the cluster $n_1$, for different values of $N=20,\ldots,5000$ with the ratio $n_1/N=1/20$ constant. Interestingly, we get a pattern similar to Fig.~\ref{figMBa}. As $N$ increases, both $M$ and $B$ increase initially but reverse the behavior after some time. Here also $B$ starts decreasing before $M$.

\item[(c)] Constant $n_1$, varying $N=n_1+n_2$, $n_1\le n_2$ ($M-B$ phase diagram not shown): Lastly, we hold $n_1=5$ constant, varying total number of nodes $N=10,\ldots,5000=n_1+n_2$. Surprisingly, the values of $B$ and $M$ for cluster $n_1$ do not change. Thus, $B$ and $M$ depend on the size of the small cluster, regardless of the size of the large cluster.
\end{itemize}

Thus, our study of $M-B$ phase diagram reveals the following features of bunching as the number of nodes is varied.
\begin{itemize}
    \item There is no bunching for a cluster of size one.
    \item As the size of the small cluster increases the bunching effect increases reaching a maximum for $n_1\sim 10$ for $m=3$. The maximum shifts to larger values of $n_1$ as $m$ increases, reaching a maximum for $n_1\sim 25$ for $m=4.5$.
    \item The increase in the total number of nodes $N$, does not affect the bunching behavior.
    \item The bunching in $n_1$ depends only on size of $n_1$, the size of large cluster $n_2$ ($\ge n_1$) does not affect bunching in $n_1$. 
\end{itemize}

\section{Different network types}

Our analysis so far indicates the importance of network structure for the bunching of EEs. Since, the walkers are independent of each other, the correlations between the EEs leading to the bunching phenomena, must solely arise due to the network structure. What type of network structures can lead to this bunching phenomena?

\begin{itemize}
\item[(i)] Other types of modular networks: We have so far considered two clusters connected by a bridge with each cluster completely connected (Network $\cal{A}$ in Fig.~\ref{fignetAB}). Let us first decrease the connections within a cluster. As an extreme, we get a tree type structure. A tree can be easily divided in two clusters by treating any edge as a bridge. We observe bunching in all these networks upto tree structure as we decrease the connections in a cluster.

Secondly, if we increase the connections by increasing the number of bridges, the bunching and correlations among EEs start decreasing, eventually losing them completely.

\item[(ii)] Random graphs: We have simulated several random graphs including network $\cal{B}$ and do not find any bunching effects.

\item[(iii)] Other types of graphs: We have considered several other types of graphs such as small world networks, scale free networks, etc. Mostly, we did not find any bunching. But, when the structure is such that a small part of the network can be treated as a cluster, sparsely connected with the remaining part we observe bunching.
\end{itemize}
\begin{figure}
    \centering
    \includegraphics[width=1\linewidth]{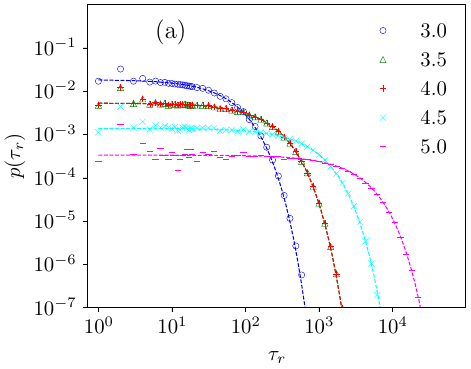}
\includegraphics[width=1\linewidth]{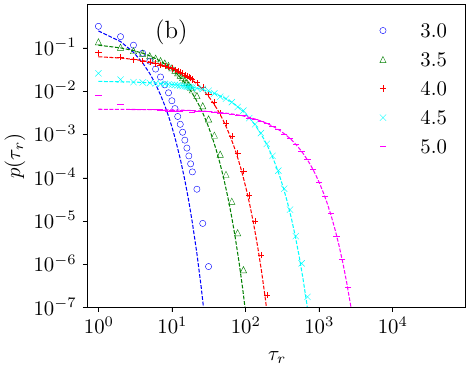}
    
    \caption{(a) and (b) The figures show the observed probability distribution $p(\tau_r)$ of recurrence time $\tau_r$ for a small cluster $n_1=5$ and a large cluster $n_2=95$, for five different values of $m$, for the Barabási–Albert scale-free network with $N=100$, preferential attachment parameter 3. The small cluster is chosen as a small part towards the edges of the scale-free structure. The dashed lines (color online) show iid values, Eq.~(\ref{ptaur}).}
    \label{fig:BA_recur}
\end{figure}
\begin{figure}
    \centering
    \includegraphics[width=1\linewidth]{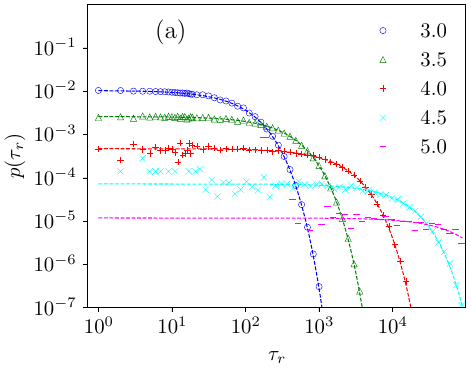}
\includegraphics[width=1\linewidth]{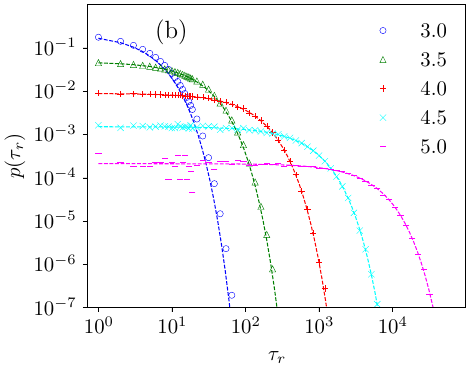}
    
    \caption{(a) and (b) The figures show the observed probability distribution $p(\tau_r)$ of recurrence time $\tau_r$ for a small cluster $n_1=5$ and a large cluster $n_2=95$, for five different values of $m$, for the Watts–Strogatz small-world network, rewired with probability $0.6$. The dashed lines (color online) show iid values, Eq.~(\ref{ptaur}).}
    \label{fig:SW_recur}
\end{figure}

\begin{figure}
    \centering
    \includegraphics[width=1\linewidth]{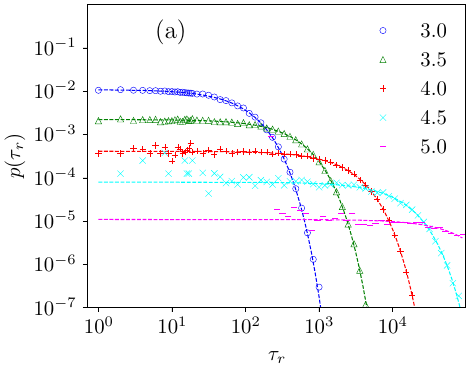}
\includegraphics[width=1\linewidth]{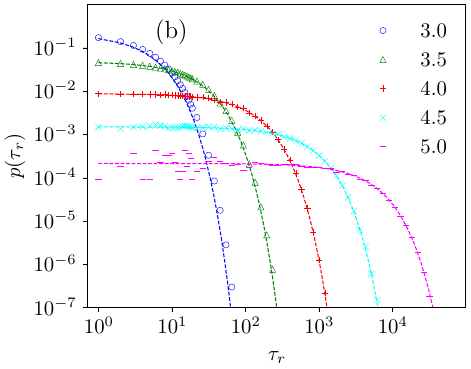}
    
    \caption{(a) and (b) The figures show the observed probability distribution $p(\tau_r)$ of recurrence time $\tau_r$ for a small cluster $n_1=5$ and a large cluster $n_2=95$, for five different values of $m$, for the Erd\H{o}s–R\'nyi random network with $N=100$, 
connection probability $p=0.6$.  The small cluster is chosen randomly. The dashed lines (color online) show iid values, Eq.~(\ref{ptaur}).}
    \label{fig:ER_recur}
\end{figure}

We now consider some networks as illustrative examples for the above observations. In Figs.~\ref{fig:BA_recur}(a) and (b) we plot recurrence time distribution $p(\tau_r)$ for a Barab\'asi–Albert scale-free network, for a small cluster $n_1=5$, and a large cluster $n_2=95$, respectively. For both clusters, the observed values match fairly well with the iid values, Eq.~(\ref{ptaur}). Thus, there are no correlation and hence no bunching of EEs. Similar point is illustrated in Figs.~\ref{fig:SW_recur}(a) and (b) for a Watts–Strogatz small-world network with $N=5+95=100$ and again there are no correlations and hence no bunching of EEs.

\begin{figure}
    \centering
    \includegraphics[width=1\linewidth]{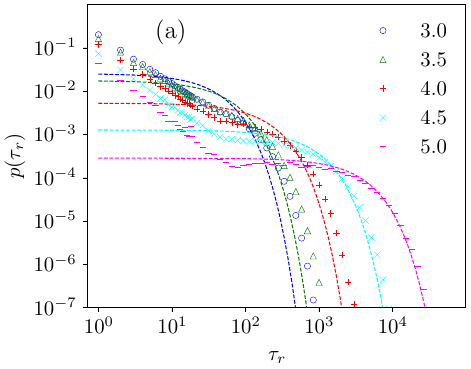}
\includegraphics[width=1\linewidth]{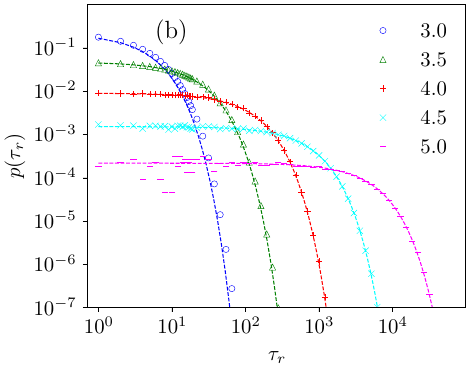}
    
    \caption{(a) and (b) The figures show the observed probability distribution $p(\tau_r)$ of recurrence time $\tau_r$ for a small cluster $n_1=5$ and a large cluster $n_2=95$, for five different values of $m$, for the Erd\H{o}s–R\'nyi random network with $N=100$, 
connection probability $p=0.6$.  The small cluster is chosen such that it is connected to the large cluster with a single bridge. The dashed lines (color online) show iid values, Eq.~(\ref{ptaur}).}
    \label{fig:ER2_recur}
\end{figure}

As our last illustration, in Figs.~\ref{fig:ER_recur}(a) and (b) we plot recurrence time distribution $p(\tau_r)$ for Erd\H{o}s–R\'enyi (ER) random network, for a small cluster and a large cluster, respectively. The small cluster is chosen randomly. The observed values match well with the iid values, showing no correlations and no bunching of EEs. On the other hand, in Figs.~\ref{fig:ER2_recur}(a) and (b) we repeat similar plots for recurrence time distribution $p(\tau_r)$ for a Erd\H{o}s–R\'enyi (ER) random network, but now the small cluster is such that it is connected to the large cluster with a single bridge. The observed values match well with the iid values for the large cluster, but the small cluster shows deviations from the iid values, thus showing correlations and bunching of EEs.

Thus, our study of different network types tells us that the requirement for bunching is a small cluster that is sparsely connected to the rest of the network.

\section{Conclusions}

In summary, our study highlights the important role of network structure in the bunching of EEs. We observe bunching in a small cluster that is sparsely connected with the rest of the network. The primary reason for this bunching is the oscillatory behavior of the probability difference $Q_{n_j}(t)$, which leads to a higher probability of EEs when $Q$ is positive and a very low probability when $Q$ is negative. The oscillatory behavior also means that the bunching is recurrent. We observe stretched exponential behavior for the recurrence time, autocorrelation function, and no. of bursty trains, indicating correlations among EEs. We note that in our model the correlations among EEs and hence the bunching emerge naturally.

Since our model is a stochastic model, and the observed values for real systems are also stochastic in nature. Thus, a direct comparison between the values obtained by our model and the observed values is not possible. However, some general conclusions can be compared. We have already noted that our model gives a bunching of EEs similar to the observed bunching of EEs. We also note that the stretched exponential behavior of the recurrence time distribution that we observe is reported for observations of several real systems \cite{PhysRevE.73.026117,PhysRevLett.94.048701}.

\section{Acknowledgments}
Financial assistance from UGC, India, and IoE scheme of MoE, India are gratefully acknowledged.
\bibliography{article}
\section{Appendix}

Here, we give numerical values of different parameters.

\begin{table}[h]
\caption{\label{tableptaur} Parameters for the stretched exponential fit \(p(\tau_r)=C_r\exp^{-D_r\tau_r^{\gamma_r}}\) for the recurrence time distribution for the small cluster $n_1=5$ (network $\cal{A}$($5-95$)) for different values of \(m\)  (Fig.~4(a)).}
\begin{tabular}{|c|c|c|c|}
%\toprule
\hline
$m$ & $C_r$ & $D_r$ &{$\gamma_r$} \\
%\midrule
\hline
3.0 & 3.401 & 2.976  & 0.236 \\
3.5 & 1.006 & 2.300  & 0.247 \\
4.0 & 0.381 & 1.923  & 0.256 \\
4.5 & 0.126 & 1.533  & 0.289 \\
5.0 & 0.050 &1.285   & 0.326 \\

%\bottomrule
\hline
\end{tabular}
\end{table}

\begin{table}[h]
\caption{\label{tableauto} Parameters for the stretched exponential fit \(p(t_d)=C_d\exp^{-D_dt_d^{\gamma_d}}\) for the Auto correlation function for the small cluster $n_1=5$ (network $\cal{A}$($5-95$)) for different values of \(m\)  (Fig.~4(b)).}
\begin{tabular}{|c|l|l|c|}
\hline
$m$ & {$C_d$} & {$D_d$} & $\gamma_d$ \\
\hline
%\midrule
3.0 & 0.167 & 0.123  & 0.787 \\
3.5 & 0.111 & 0.205  & 0.686 \\
4.0 & 0.078 & 0.365  & 0.573 \\
4.5 & 0.086 & 1.026  & 0.367 \\
5.0 & 1.035 & 3.901  & 0.168 \\

\hline
%\bottomrule
\end{tabular}
\end{table}

\begin{table}[h]
\caption{\label{tablebsizedist} Parameters for the stretched exponential fit \(p_{\Delta t}(b)=C_b\exp^{-D_bb^{\gamma_b}}\) for the  burst size distribution for the small cluster $n_1=5$ (network $\cal{A}$($5-95$)) for different values of \(\Delta t\)  (Fig.~5(b)).}
\begin{tabular}{|r|r|r|r|}
%\toprule
\hline
{$\Delta t$} & {$C_b$} &{$D_b$} & {$\gamma_b$} \\
%\midrule
\hline

4  & 22334.721 &10.125& 0.305 \\
16 & 736.378   & 6.868& 0.309 \\
64 &9.422      &2.727 & 0.494 \\
256&0.671 & 0.689 & 0.769 \\

%\bottomrule
\hline
\end{tabular}
\end{table}

\begin{table}[h]
\caption{Observed mean recurrence time and standard deviation, $\langle \tau_{r} \rangle$  and $\sigma_{r}$, and the theoretical (i.e. for iid EEs) values, $\langle \tau_r^{iid} \rangle $ and $\sigma_{r}^{iid}$, for the small and big clusters (network $\cal{A}$($5-95$)) for different values of $m$. \label{tablemstaur}}
\renewcommand{\arraystretch}{1.4}  % Increase row height slightly for better readability
\setlength{\tabcolsep}{1pt}  % Adjust column separation to avoid cutting off
\begin{tabular}{|c|r|r|r|r|}
\hline
\multicolumn{5}{|c|}{\textbf{For the Small Cluster ($n_1=5$)}} \\
\hline
$m$ &$\langle {\tau_{r}} \rangle $ & $\sigma_{r}$ & $\langle {\tau_{r}^{iid}} \rangle$ & $\sigma_{r}^{iid}$ \\
\hline
3.0 &26.380   &48.185   &26.378   &25.874 \\
3.5 &74.865    &119.995  &74.946   &74.444 \\
4.0 &230.130   &319.029  &230.206  &229.705 \\
4.5 &862.052  &1039.630 &862.919  &862.419 \\
5.0 &3089.245  &3411.514 &3091.691 &3091.192 \\
\hline
\end{tabular}

\vspace{0.5cm} % Add vertical space between sections

\begin{tabular}{|c|r|r|r|r|}
\hline
\multicolumn{5}{|c|}{\textbf{For the Big Cluster ($n_2=95$)}} \\
\hline
$m$ & $\langle {\tau_{r} \rangle}$ & $\sigma_{r}$ & $\langle {\tau_{r}^{iid}}\rangle$ & $\sigma_{r}^{iid}$ \\
\hline
3.0 &4.672   &4.631    & 4.671    &4.141 \\
3.5 &22.627  &22.556   & 22.627   &22.121 \\
4.0 &131.955 &131.988  & 132.048  &131.547 \\
4.5 &920.980 & 920.047 & 921.685  &921.185 \\
5.0 &7623.120& 7646.326& 7658.635 & 7658.135 \\
\hline
\end{tabular}
\end{table}

\end{document}